# Tailoring giant quantum transport anisotropy in disordered nanoporous graphenes

Isaac Alcón[1]*, Aron Cummings[1]* and Stephan Roche[1,2]

[1]Catalan Institute of Nanoscience and Nanotechnology (ICN2), CSIC and BIST, Campus UAB, Bellaterra, 08193 Barcelona, Spain

[2]ICREA, Institució Catalana de Recerca i Estudis Avançats, 08070 Barcelona, Spain

*Corresponding authors: isaac.alcon@icn2.cat, aron.cummings@icn2.cat

## Abstract

During the last 15 years bottom-up on-surface synthesis has been demonstrated as an efficient way to synthesize carbon nanostructures with atomic precision, opening the door to unprecedented electronic control at the nanoscale. Nanoporous graphenes (NPGs) fabricated as two-dimensional arrays of graphene nanoribbons (GNRs) represent one of the key recent breakthroughs in the field. NPGs interestingly display in-plane transport anisotropy of charge carriers, and such anisotropy was shown to be tunable by modulating quantum interference. Herein, using large-scale quantum transport simulations, we show that electrical anisotropy in NPGs is not only resilient to disorder but can further be massively enhanced by its presence. This outcome paves the way to systematic engineering of quantum transport in NPGs as a novel concept for efficient quantum devices and architectures.

## Introduction

Bottom-up on-surface synthesis is a powerful tool to fabricate carbon nanostructures with atomic precision.[1–3] During the last 15 years, this method has allowed the synthesis of different types of nanomaterials such as 2D covalent organic frameworks,[4,5] magnetic nanographenes,[6,7] 1D π-conjugated polymers,[8,9] and graphene nanoribbons (GNRs),[2,10] among others.[4,11] GNRs, concretely, have gathered great attention for carbon nanoelectronics,[12,13] given their semiconducting electronic structure and chemical stability which has facilitated their integration in various solid-state devices.[14–16] The electronic structure of GNRs has also been shown to be highly tunable by bottom-up design, permitting the realization of topologically non-trivial states,[17,18] which has created high expectations for their use in quantum technologies.[19,20]

Another related family of carbon nanomaterial which could strongly benefit from the key features of GNRs is the family of nanoporous graphenes (NPGs). NPGs, first reported in 2018,[21] are formed as 2D arrays of GNRs laterally connected via π-conjugated covalent bonds, thus creating a regular array of pores in the otherwise graphene-like 2D structure. Since the pioneering work, different types of NPGs have been reported,[22,23] and have been integrated in solid-state devices exploiting their semiconducting characteristics.[21,24] An important property of the NPGs is their in-plane charge transport anisotropy. It was theoretically shown that current preferentially flows along the GNRs, but due to their electronic coupling it also spreads in the transversal direction.[25] Importantly, due to the fully π-conjugated nature of NPGs, such inter-ribbon coupling may be tuned by bridging the GNRs with π-conjugated molecular bridges. For example, single phenyl rings determine the strength of inter-ribbon transport depending on their connectivity,[26] due to destructive quantum interference (QI).[27] Para connections lead to transversal spreading in the so-called para-NPG material (as in the original NPG[25]), whereas meta connections cut inter-ribbon coupling due to QI, leading to almost complete 1D transport in the resulting meta-NPG.[26] Importantly, NPGs including para- and meta- connections between GNRs have recently been synthesized,[28] highlighting the experimental feasibility of such QI-engineering of NPGs.

Despite the great promise of these strategies to realize highly tunable platforms based on NPGs for multiple applications, it remains unclear whether QI-engineering is robust under realistic material and device conditions. Indeed, to date all theoretical transport studies have exclusively considered the pristine (ultra-clean) systems,[25,26,29] without evaluating how static or dynamic disorder could influence NPGs electrical characteristics. In particular, electrostatic disorder, which may arise from interactions with underlying substrates or via chemical/atomic adsorbates,[30] could strongly influence the effectiveness of QI-engineering.[26,28] More generally, a priori it is difficult to predict the effect of local impurity scattering on the transport along each in-plane direction in NPGs. Therefore, evaluating the impact of disorder on electrical transport in NPGs is of primary importance to discern their future applicability in various emerging fields such as nanoelectronics, nanosensing, spintronics and quantum technologies.

Here we evaluate, via efficient large-scale quantum transport simulations, the effect of electrostatic disorder on the transport characteristics of three different types of NPGs: the original fabricated NPG, the para-NPG, and the meta-NPG. Our quantum simulations are based on an efficient wave packet propagation method and the use of the Kubo formalism that gives access to the length- and energy-dependent-conductivity of disordered systems containing many millions atoms.[31] Electrostatic disorder is modelled as local electron-hole puddles randomly distributed throughout each sample. By simulating the time evolution of wave packets, we obtain fundamental insight into the transport dynamics, and we characterize the transport regime for each material and in-plane direction. Altogether, we find that the electrical anisotropy in NPGs is strongly enhanced by the presence of electrostatic disorder. This derives from the stronger effect of charged impurities on transport between GNRs than along GNRs. While this is true for all considered NPGs and should be general for this class of 2D materials, the transport anisotropy is maximum for the meta-NPG case. In this material transport through the bridges is much more sensitive to disorder than in the other NPGs, making it fully insulating in one in-plane direction, while remaining a semiconductor in the other direction (along the GNRs). Consequently, to the best of our knowledge, meta-NPG is the first 2D material displaying an unprecedented giant quantum transport anisotropy, reaching the limit of behaving as a 2D array of purely isolated 1D nanoelectronic channels conveying charges in parallel. Overall, our results indicate that the use of NPGs to design circuit architectures based on guided transport and controlled switching through 1D channels is feasible under realistic experimental conditions.

## Results and discussion

**Pristine systems**

Fig. 1a depicts the atomic structure of our considered materials which, with graphene as a reference, include the original NPG[21] and the subsequently predicted and chemically tailored para-NPG and meta-NPG.[26] Structurally, NPGs may be thought of as a 2D grid of parallel GNRs laterally connected via strong covalent bonding. While in NPG the GNRs are directly linked via single C-C bonds, in para- and meta-NPGs the GNRs are bridged via a phenylene ring in para- and meta-configuration, respectively (Fig. 1a-top). As previously shown,[25,26] the basic electronic structure of NPGs is properly captured with a standard first nearest-neighbour tight-binding (TB) Hamiltonian, as normally used for graphene:

Eq. (1) $$\hat{H} = t \sum_{\langle i,j \rangle} \hat{c}_i^\dagger \hat{c}_j,$$

where $\hat{c}_i^\dagger$ ($\hat{c}_j$) is the creation (annihilation) operator for $p_z$ orbitals at site $i$ ($j$), $t$ is the hopping parameter (2.7 eV), and the sum runs over first nearest neighbours. Fig. 1a displays the resulting band structures for each material. In contrast to graphene, all NPGs are semiconductors with a sizeable band gap of around 0.6 eV. Each NPG displays two conduction (valence) bands contributing to charge transport near the Fermi level ($E_F$). It has been shown that the momentum splitting of these bands, at a given energy, is directly proportional to the inter-ribbon electronic coupling.[25,26] Thus, qualitatively, the inter-ribbon coupling progressively decreases from NPG towards the meta-NPG, where the two conduction (valence) bands are nearly degenerate for the entire 2 eV range around $E_F$.

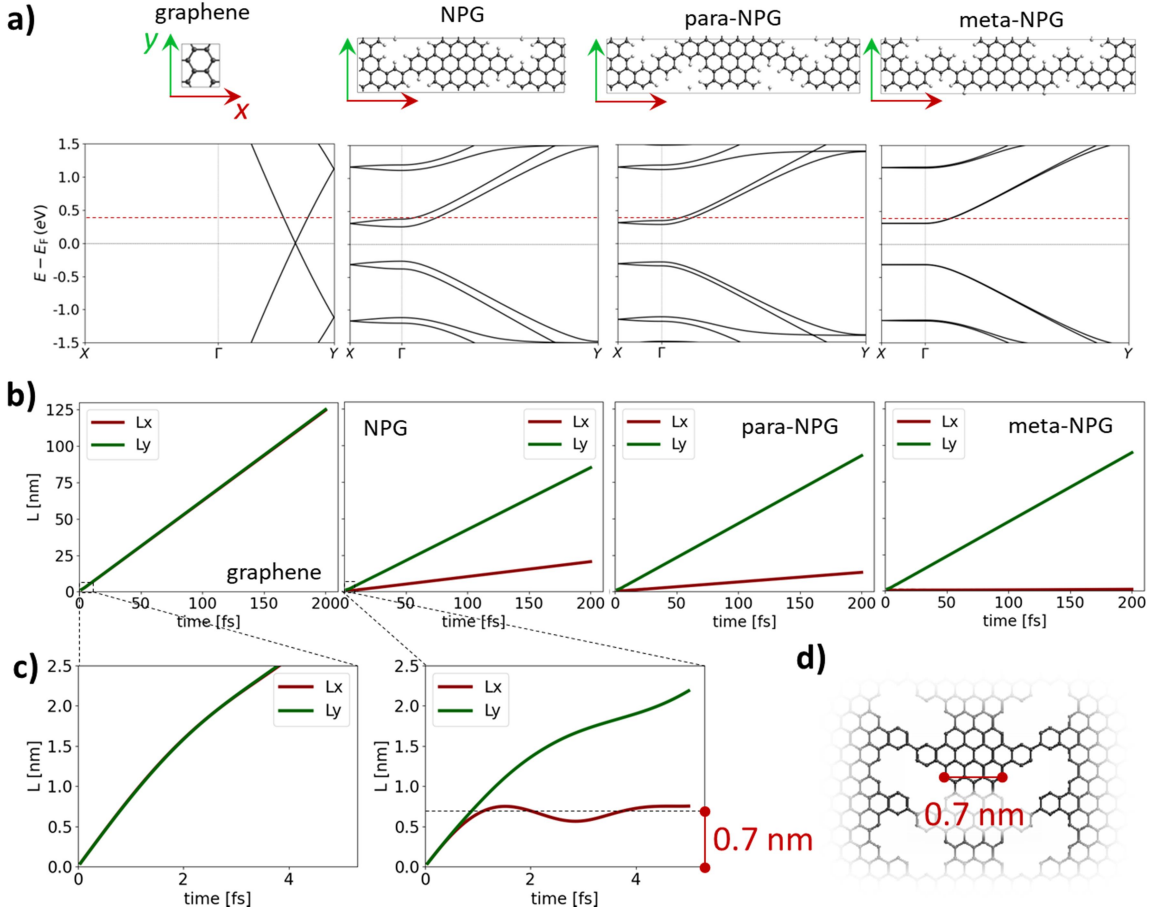

Fig. 1. a) Periodic structure of each material (top) and associated band structure (bottom) obtained from the 1$^{st}$ nearest-neighbour TB model ($t_{ij}$ = -2.7 eV, $\varepsilon_0$ = 0 eV). As highlighted with coloured arrows, the $y$ ($x$) axis is parallel (perpendicular) to the GNRs in NPGs. The red dashed line indicates the energy at which the wave packet propagation is calculated ($E − E_F$ = 0.4 eV). b) Wave packet displacement along each in-plane direction ($L_x$: red curves; $L_y$: green curves). c) Wave packet propagation lengths for the first 5 fs of the run in graphene and the NPG. The black dotted line indicates the average width of the GNR composing all NPGs, as depicted in d).

To evaluate the effect of disorder on the quantum transport properties of NPGs, we make use of a real-space order-N wave packet propagation methodology, generally known as the linear-scaling quantum transport (LSQT) method, pioneered decades ago for the study of aperiodic and disordered materials.[32–34] LSQT is based on the Kubo formalism and allows the simulation of quantum transport in complex disordered systems composed of many millions of atoms. Its various implementations have been recently presented in Ref. [31]. The central quantity of this method is the energy- and time-dependent mean squared displacement, which is here computed in the two in-plane directions,

Eq. (2) $$\Delta X^2(E,t) = \frac{Tr[\delta(E-\hat{H})|\hat{X}(t)-\hat{X}(0)|^2]}{\rho(E)},$$

Eq. (3) $$\Delta Y^2(E,t) = \frac{Tr[\delta(E-\hat{H})|\hat{Y}(t)-\hat{Y}(0)|^2]}{\rho(E)},$$

where $\hat{Y}(t)$ and $\hat{X}(t)$ are the position operators along (y) and perpendicular to (x) the GNRs direction, respectively, and $\rho(E) = Tr[\delta(E - \hat{H})]$ is the density of states of the system at a given energy. We then obtain the propagation length along each direction as

Eq. (4) $$L_x = \sqrt{\Delta X^2},$$

Eq. (5) $$L_y = \sqrt{\Delta Y^2}.$$

We first focus on the pristine NPG systems. We build ca. 300x300 nm² samples for each material, composed of approximately three million atoms (see Table 1).

Table 1. Sample dimensions: Number of unit cell repetitions along *x* ($n_x$) and *y* ($n_y$) directions, large-scale box dimensions along *x* (A) and *y* (B) directions and total number of atoms ($N_{at}$).

|  | graphene | NPG | para-NPG | meta-NPG |
|---|---|---|---|---|
| $n_x$ ; $n_y$ | 174/200 | 92/345 | 75/345 | 75/345 |
| A ; B [nm] | 300.6/299.3 | 301.6/299.9 | 301.9/299.9 | 301.9/299.9 |
| $N_{at}$ | 3,340,800 | 2,539,200 | 2,380,500 | 2,380,500 |

Due to the electron-hole symmetry present in all considered systems (see band structures in Fig. 1a) we limit our discussion to energies lying in the conduction bands, with the resulting conclusions also applicable to the valence bands. We consider quantum transport at $E - E_F$ = 0.4 eV (see dashed red line in Fig. 1a), while in the supporting information (SI) we show qualitatively similar results for $E - E_F$ = 0.7 eV.

Fig. 1b shows the evolution of $L_y(t)$ (green) and $L_x(t)$ (red) for each pristine material. While for graphene the evolution of $L_y(t)$ and $L_x(t)$ exactly overlap, for the NPGs $L_x(t)$ is always significantly smaller than $L_y(t)$. Zooming in on the first 5 fs (Fig. 1c) we find that this divergence takes place around the first femtosecond when the wave packet reaches 0.7 nm, corresponding to the average width of the GNRs, as highlighted in Fig. 1d. This ultra-fast process of hitting the GNR wall is common to all NPGs (see Fig. S1 in SI). However, while the $L_y(t)$ evolution is similar for all NPGs, $L_x(t)$ monotonically decreases from NPG to meta-NPG (Fig. 1b), where $L_x(t)$ appears as a nearly flat curve. This reflects the fact that while para-connected phenylene rings slightly decrease inter-ribbon transmission as compared to direct C-C bonding (para-NPG vs NPG), the meta-configuration almost entirely suppresses inter-ribbon transport. This difference between NPGs, not noticeable for very short times (Fig. S1), becomes clear around 10-20 fs (Fig. S2).

To quantify the electrical anisotropy of the pristine systems, we first extract the Fermi velocity along each direction as

Eq. (6) $$v_i = \frac{dL_i}{dt},$$

where *i* denotes the *x* or *y* direction. To avoid the initial fluctuations of $v_i$, we average over the second half of each run (see Methods for details) to obtain the electrical anisotropy (A) as the ratio between $\bar{v}_y$ and $\bar{v}_x$. Table 2 shows the resulting anisotropy values for each material.

Table 2. Transport anisotropy at E – $E_F$ = 0.4 eV for each material in the pristine form.

|  | graphene | NPG | para-NPG | meta-NPG |
|---|---|---|---|---|
| $A = \overline{v}_y/\overline{v}_x$ | 1.0 | 4.1 | 7.1 | 120.0 |

These values agree with the qualitative picture previously obtained from $L_y(t)$ and $L_x(t)$ (Fig. 1b) and are fundamentally the same for states deeper within the conduction band (see Table S1 for E – $E_F$ = 0.7 eV). These anisotropy values are also in line with previous predictions for other phenylated NPGs which were recently experimentally fabricated,[28] and are of the same order of magnitude as other low-symmetry 2D materials proposed for in-plane anisotropic electronics,[35] including black phosphorous.[36] However, it is worth highlighting the special case of meta-NPG (see Table 2 and S1) where the anisotropy is at least one order of magnitude larger than the other NPGs and most low-symmetry 2D materials.[35]

**Disordered systems**

It is known that electrostatic disorder may severely affect transport anisotropy in 2D materials, up to the point of reversing the preferential direction for current flow.[36] Such disorder may arise in experiments from interaction with the underlying substrate,[30] or due to molecular/atomic adsorbates.[37] Therefore, it is imperative to evaluate whether the high anisotropy of NPGs survives under these conditions. Electrostatic disorder may be effectively modelled via a random distribution of Gaussian-like electron-hole puddles. These are included in the TB model by modifying the on-site terms as

Eq. (7) $$\varepsilon_i = \sum_j V_j exp\left(-|\vec{r}_i - \vec{r}_j|^2/2\beta^2\right),$$

where $\vec{r}_i$ is the position of each carbon site, $\vec{r}_j$ is the position of each puddle centre, $\beta$ is the puddle width set to 4.35 Å, and $V_j$ is the potential height, which is randomly distributed between -2.8 and +2.8 eV (see Methods for further details). We use a puddle density of 0.1% which, as shown in previous studies,[38] leads to charge mobility values in graphene similar to experiments.[37] For illustration, a puddle distribution obtained with these parameters is shown in Fig. 2a for one-fourth of the total area of our NPG sample (see square outline at bottom left in Fig. 2a). There we see that the puddles have an average diameter similar to the GNR width.

Under such disorder, transport in 2D materials evolves from the ballistic regime, characteristic of pristine systems, to the diffusive or even localization regimes depending on the system length and disorder strength.[31] The transport regime can be actually extracted from the scaling of the diffusion coefficient, which is obtained from the time evolution of the mean squared displacement as

Eq. (8) $$D_x(E,t) = \frac{1}{2}\frac{\partial}{\partial t}\Delta X^2(E,t),$$

Eq. (9) $$D_y(E,t) = \frac{1}{2}\frac{\partial}{\partial t}\Delta Y^2(E,t).$$

This, in turn, allows us to extract the associated time- and energy-dependent electrical conductivity for each in-plane direction as

Eq. (10) $$\sigma_x(E,t) = e^2\rho(E)D_x(E,t),$$

Eq. (11) $$\sigma_y(E,t) = e^2 \rho(E) D_y(E,t),$$

where $e$ is the elementary charge. Due to the time evolution of the wave packet lengths ($L_x$ and $L_y$) we finally obtain a length-dependent conductivity, at a given energy, for each in-plane direction: $\sigma_x(L_x)$ and $\sigma_y(L_y)$.

Fig. 2 shows $\sigma_x(L_x)$ and $\sigma_y(L_y)$ for each material under the influence of electrostatic disorder. We consider 10 random puddle distributions to obtain statistically averaged quantities (see Methods for details). In Fig. 2b, graphene undergoes a fast transition from a ballistic transport regime to a diffusive one, where conductivity remains approximately constant with increasing $L$. Additionally, there is no significant difference between $\sigma_x$ and $\sigma_y$, indicating the isotropic nature of this material. This picture changes dramatically for NPGs (Fig. 2c-e) where there is a substantial difference in transport along each in-plane direction, with $\sigma_x \ll \sigma_y$. If we focus on transport along GNRs ($\sigma_y$; green curves), we see that the evolution of conductivity is also very different as compared to graphene. All NPGs (Fig. 2c-e) present a maximum in $\sigma_y$, known as the semi-classical conductivity ($\sigma_y^{SC}$; see arrows), followed by a continuous decay of $\sigma_y$ with increasing $L_y$. As explained in detail in Section S2 of the SI, this trend may be ascribed to the strong localization regime of transport. In this regime, quantum interference caused by random scattering events[31] eventually leads to full confinement of charge carriers (i.e. $\sigma \approx 0$). Experimentally, this corresponds to an exponential increase in resistance with device size. Looking at $\sigma_x$ (red curves), one can note the strong reduction of transport in this in-plane direction as compared to $\sigma_y$ for all NPGs (Fig. 2c-e). As shown in the bottom panel of Fig. 2c (NPG), $\sigma_x$ follows a similar trend as $\sigma_y$ but with a much lower magnitude. This decrease in $\sigma_x$ is even more pronounced for para-NPG (bottom panel in Fig. 2d), highlighting the less conductive character of para-connected phenyl rings as compared to direct C-C bonding. Finally, in the meta-NPG (Fig. 2e) $\sigma_x$ becomes completely suppressed, reaching the situation where $\sigma_x \approx 0$ (see inset in bottom panel of Fig. 2e). This result demonstrates the efficacy of QI to effectively cut transport between GNRs, even under the presence of strong electrostatic disorder. As with $\sigma_y$, the decay rate of $\sigma_x$ for all three NPGs may be associated with the strong localization regime (see Section S2 in the SI and further discussion below). It is also worth noting that upon reducing transport between GNRs (i.e. $\sigma_x^{metaNPG} < \sigma_x^{paraNPG} < \sigma_x^{NPG}$), transport along the GNRs also becomes increasingly degraded with disorder (i.e. $\sigma_y^{metaNPG} < \sigma_y^{paraNPG} < \sigma_y^{NPG}$). In other words, the more 1D-like the electronic system is, the more sensitive it becomes to disorder, which may be explained by the higher probability of back-scattering.[39] This also explains the difference in $\sigma_y$ between all NPGs and graphene where, due to its purely 2D transport nature, localization effects are much weaker (Fig. 2b).

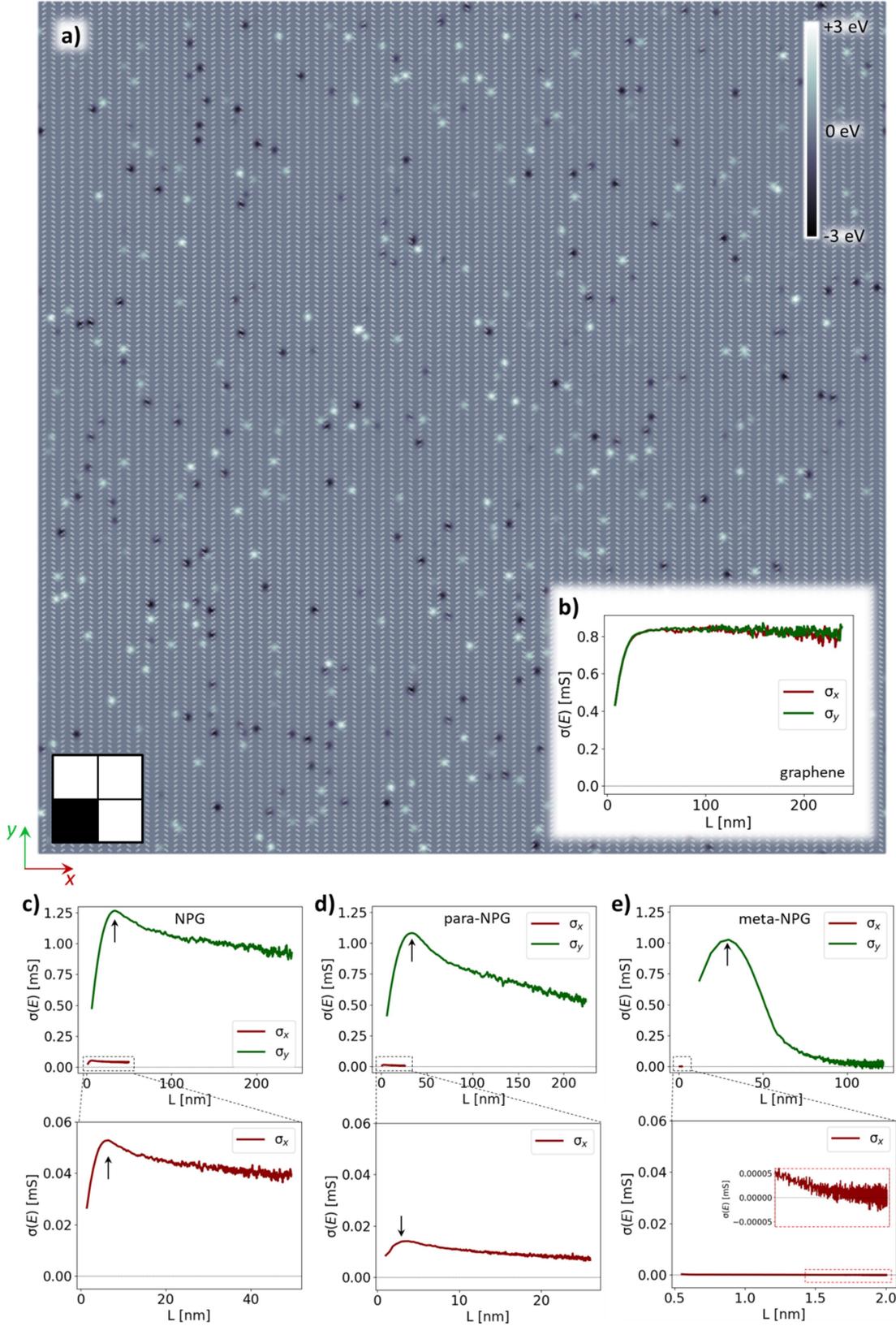

Fig. 2. a) Puddle distribution map for one-fourth of the total area (300x300 nm²) of the NPG sample (see square outline at bottom left). Bright and dark regions are associated with p-doping ($\varepsilon_i > 0$) and n-doping ($\varepsilon_i < 0$), respectively. Evolution of $\sigma_y$ (green curves) and $\sigma_x$ (red curves) with wave packet length at E – E$_F$ = 0.4 eV for b) graphene, c) NPG, d) para-NPG and e) meta-NPG. For the last three cases, an additional $\sigma_x(L_x)$ plot is provided at the bottom for better visualization and comparison.

Defining the charge transport anisotropy in such a complex scenario is not straightforward. One way is to consider the ratio between semi-classical values of conductivity along each in-plane direction,

Eq. (12) $$A = \frac{\sigma_y^{sc}}{\sigma_x^{sc}}.$$

$\sigma_y^{sc}$ ($\sigma_x^{sc}$) can be extracted as the maxima in the $\sigma_y(L_y)$ ($\sigma_x(L_x)$) curves (see arrows in Fig. 2c-e), and represent an upper estimate for transport in each material under such disorder conditions. Such an approximation may be valid for devices working at high temperature, where electron-phonon coupling can wash out the quantum interference effects that lead to strong localization. Such dephasing effects may increase conductivity at all lengths up to the semi-classical value,[39] justifying the use of $\sigma^{sc}$ to characterize transport and electrical anisotropy at moderate to high temperatures.

Table 3 provides the resulting anisotropy values. While graphene's characteristic isotropic transport remains unchanged, both NPG and para-NPG become more anisotropic than in their pristine form (see Table 2 for comparison). This indicates that disorder perturbs transport more effectively between GNRs than along GNRs. This may be quantified via the ratio of momentum relaxation times which may be extracted from the relation $\sigma_y^{sc}/\sigma_x^{sc} = (v_y/v_x)^2 \cdot \tau_y/\tau_x$. We obtain $\tau_y/\tau_x$ values of 1.4 and 1.5 for the NPG and para-NPG, respectively.

Table 3. Electrical transport anisotropy at E − E$_F$ = 0.4 eV for each material under the presence of disorder.

|  | graphene | NPG | para-NPG | meta-NPG* |
|---|---|---|---|---|
| $A = \dfrac{\sigma_y^{sc}}{\sigma_x^{sc}}$ | 1.0 | 23.7 | 76.7 | ∞ |

*Note: ∞ reflects the fact that meta-NPG does not display a diffusive peak for $\sigma_x(L_x)$ and so qualitatively $\sigma_x^{sc} \approx 0$. If one calculates anisotropy A as $\sigma_y^{sc}/\sigma_x^{max}$, then A = 3781.3. However, this value is not representative of the actual anisotropy in this system, as $\sigma_x^{max}$ here is associated with the initial ultra-fast ballistic regime that quickly evolves to full confinement where $\sigma_x(L_x) = 0$ (see inset in bottom panel of Fig. 2e).

Meanwhile, the meta-NPG exhibits unique behavior. In this material there is effectively no diffusive maximum because transport directly transits from "ballistic spreading" at short times to a fully confined situation, as may be seen in the inset of Fig. 2e (bottom panel). Therefore, one may argue that $\sigma_x^{sc} \approx 0$ and, due to the finite $\sigma_y^{sc}$ (see arrow in Fig. 2e, green curve) one could conclude that $A \approx \infty$ for this 2D material. This peculiar result indicates that meta-NPG behaves as a semiconductor along the GNRs, but as an electrical insulator in the perpendicular direction.

**meta-NPG: A host for giant quantum transport anisotropy**

Although stating that $A \approx \infty$ for meta-NPG may appear excessive, there is an alternative way to look at the problem at hand that supports this conclusion. Fig. 3 shows the comparison between the wave packet propagation lengths perpendicular to the GNR ($L_x$ in 3a) and parallel to them ($L_y$ in 3b) for different puddle distributions (light curves) and the resulting average (darker curves). As expected from $\sigma_x(L_x)$ (red curve in Fig. 2e), we see that $L_x$ saturates to a length of approximately 2 nm. This distance exactly coincides with the distance between two phenylene rings bound to a single GNR (see the unit cell at the top of Fig. 3c). It is also worth noting that all puddle distributions converge to approximately the same value of 2 nm. At E − $E_F$ = 0.7 eV $L_x$ saturates to 3 nm (Fig. S7) meaning that deeper in the conduction band charge carriers manage to reach the first neighbouring GNR, but do not propagate beyond that limit.

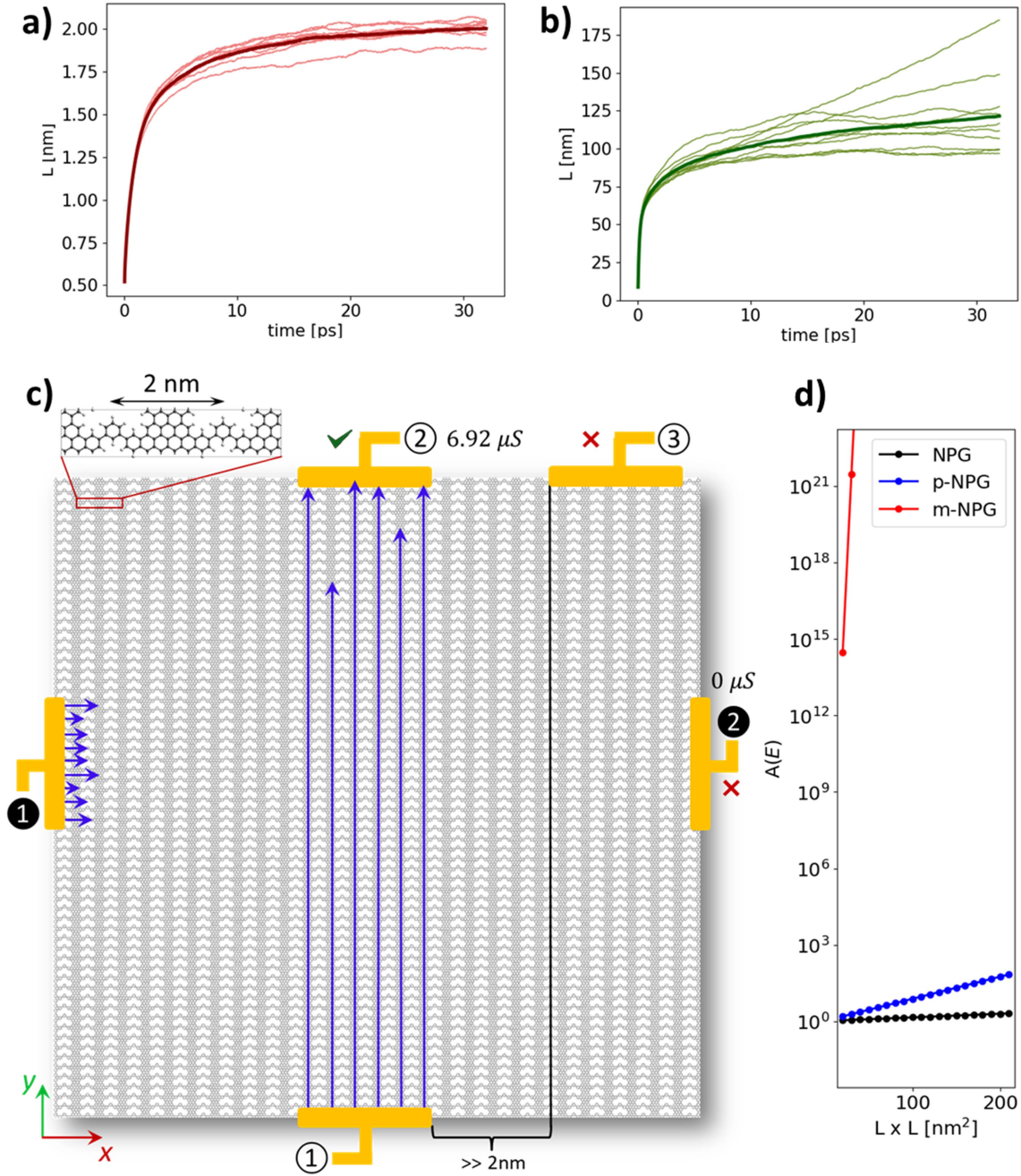

Fig. 3. Time evolution of wave packet propagation lengths for meta-NPG at E − $E_F$ = 0.4 eV for different puddle distributions (light curves) and their average (dark curves) along a) the direction perpendicular to the GNRs ($L_x$) and b) parallel to them ($L_y$). c) Schematic of a 50x50 nm² meta-NPG device displaying the expected electrical output for different electrode setups measuring transport along the GNRs (①→②) and perpendicular to them (❶→❷) including extrapolated conductance values (see Section S2 in SI). d) Transport anisotropy (log-scale) at E − $E_F$ = 0.4 eV with square device length (i.e. LxL nm²) for NPG

(black), the para-NPG (blue) and the meta-NPG (red). Anisotropy is calculated as the ratio between the extrapolated conductance values along each in-plane direction ($A = G_y/G_x$). See Eq. (13) and Section S2 for details.

On the contrary, $L_y$ grows beyond 100 nm (Fig. 3b) and, additionally, different puddle distributions lead to dispersive saturation lengths ranging from 80 nm to a few hundred nanometers. Such variability between different puddle distributions for transport along GNRs (Fig. 3b) is not present for the other NPGs (Fig. S8), which highlights that this is a unique feature of meta-NPG arising from its quasi-1D electrical nature (see Section S3 for further details and discussion).

This significant difference in saturation length between the two in-plane directions should entirely determine the behaviour of meta-NPG-based devices. If one were to measure transport in a 50x50 nm² meta-NPG device, as depicted in Fig. 3c, a finite electrical signal should be detected between electrodes ① and ②, since the average puddle propagation along the GNRs, $L_y$, grows beyond 100 nm (Fig. 3b). In the strong localization regime this may be estimated by fitting the exponentially decaying length-dependent conductance ($G_y(L_y)$ and $G_x(L_x)$),

Eq. (13) $$G(E,L) \propto exp[-L/\xi(E)],$$

where $\xi(E)$ is the localization length, related to the saturation lengths from Fig. 3a,b just discussed (see Section S2 and Table S2 for details). Eq. (13) allows us to extrapolate *G* for a particular device length, leading to $G_y(50\ nm) = 6.92\ \mu S$ (see Section S2 for full discussion and results). On the contrary, $G_x(50\ nm) = 2.5 \cdot 10^{-3}\ \mu S \approx 0\ \mu S$, and so no measurable current should be expected between ❶ and ❷ (Fig. 3c), unveiling the giant transport anisotropy existing in this meta-NPG device ($A = G_y^{50nm}/G_x^{50nm} \approx \infty$). Intuitively, this should not be restricted only for our chosen device configuration, but for any situation where the distance between ❶ and ❷ is larger than 5-10 nm since, as shown in Fig. 3a, $L_x$ is fully saturated at 2 nm (3 nm at E − E_F = 0.7 eV; see Fig. S7). This fundamentally disruptive result ($A \approx \infty$) is unique to meta-NPG, because the other NPGs have finite conductance values along both in-plane directions at similar lengths (see Table S3 in SI), leading to relatively moderate/low anisotropy values for device sizes where meta-NPG already displays giant anisotropy (see Fig. 3d). It is also worth noting that finite temperatures should more effectively wash out long-range localization effects than short-range effects.[39] Therefore, thermal dephasing should improve transport along the GNRs (i.e. higher $G_y(L)$ values) rather than improving transport perpendicular to them ($G_x(L)$) which, if anything, would even further increase anisotropy. Again, this scenario only applies to meta-NPG, where the main interference effect suppressing charge transport between GNRs occurs within single (meta-connected) phenyl rings composed of six carbon atoms. Finally we note that transport along *y* would only be measurable when the same set of GNRs (or first neighbouring ribbons) are contacted (e.g. ① to ② in Fig. 3c). Therefore, no current should be measurable between ① and ③ since the transverse distance between these two electrodes is significantly higher than the $L_x$ saturation value of 2 nm (see Fig. 3c).

**Conclusions**

We have evaluated unprecedented charge transport anisotropy of NPGs in the presence of disorder. Our results demonstrate that anisotropy survives disorder and is even strongly enhanced for all NPGs, which implies the technological applicability of this appealing feature (such as for guided transport or directional switching). However, we also find that the conductivity of these materials is significantly degraded due to the presence of electrostatic disorder. Our simulations demonstrate that the considered materials can be classified according to increasing 1D electronic character: graphene → NPG → para-NPG → meta-NPG. Such 1D nature is the main cause for the overall degradation of transport, as compared to graphene, in full accordance with transport theory of 1D systems.[39] We also find that chemically-driven bridge engineering -- which leads to quantum interference (QI) engineering -- is a powerful tool to tailor anisotropy in these materials under experimentally realistic conditions. While para-NPG behaves as a semiconductor along both in-plane directions, meta-NPG displays conducting character only along the GNRs, and is an insulator in the perpendicular direction. This is demonstrated by the full confinement of wave packets in that direction, whose propagation is entirely saturated after 2 to 3 nm, which is approximately the distance between a pair of GNRs. These results thus highlight meta-NPG as an exemplary 2D material featuring giant transport anisotropy ($A \approx \infty$).

All in all, this work highlights the potential of para- and meta-connected phenyl rings as tools to chemically engineer the transport properties of carbon nanostructures. Indeed, these concepts should be applicable to other carbon nanomaterials such as 2D covalent organic frameworks, nanographenes, 1D π-conjugated polymers, and GNR heterojunctions. More generally, our results also demonstrate the power of inheriting well-established ideas from single-molecule electronics as a guideline to design carbon nanostructures with desired functionality. We believe such a multidisciplinary approach will play a central role in the future realization of carbon-based nanoelectronics and quantum technologies.

**Methods**

The unit cells shown in Fig. 1 were used to construct first nearest-neighbour TB models with parameters generally utilized for graphene ($t_{ij}$ = -2.7 eV, $\varepsilon_0$ = 0 eV). The band structures for each material were generated with the Python-based SISL utility.[40] Unit cells were repeated along the *x* and *y* directions to obtain 300x300 nm$^2$ samples for each material (see Table 1). LSQT simulations are based on the Chebyshev polynomial expansion,[31] which in our case used 4000 moments and an energy broadening of 0.02 eV. For the pristine systems, each material was calculated using ten different initial random phases, thus obtaining statistically averaged quantities. Different time steps were used for the different plots of Fig. 1, ranging from 0.05 to 2 fs.

For the samples including disorder, we considered ten random puddle distributions of the form shown in Eq. 7. In order to speed up the Hamiltonian construction, we imposed a cutoff of the Gaussian tail at 6$\beta$ (Eq. 7), thus neglecting on-site modifications, $\varepsilon_i$, lower than 4·10$^{-8}$ eV. For each puddle distribution we ran ten initial random phase distributions, thus effectively averaging over 100 LSQT runs per material, which allowed us to significantly reduce the noise of the different calculated quantities.

**Acknowledgements**

I.A. is grateful for a Juan de la Cierva postdoctoral grant (FJC2019-038971-I) from the Ministerio de Ciencia e Innovación and acknowledges access to supercomputer resources as provided through a grant (TEMP-2023-1-0028) from the Red Española de Supercomputación (RES). ICN2 is funded by the CERCA Programme from Generalitat de Catalunya, has been supported by the Severo Ochoa Centres of Excellence programme [SEV-2017-0706] and is currently supported by the Severo Ochoa Centres of Excellence programme, Grant CEX2021-001214-S, both funded by MCIN/AEI/10.13039.501100011033.

**Conflicts of interest**

There are no conflicts of interest to declare.

We demonstrate that quantum transport anisotropy of nanoporous graphenes not only survives, but is enhanced, under the presence of electrostatic disorder. Additionally, we find that tailoring nanoporous graphenes via quantum interference effects permits realizing giant quantum transport anisotropy in these unique 2D materials.

Isaac Alcón*, Aron Cummings*, Stephan Roche

Tailoring giant quantum transport anisotropy in disordered nanoporous graphenes

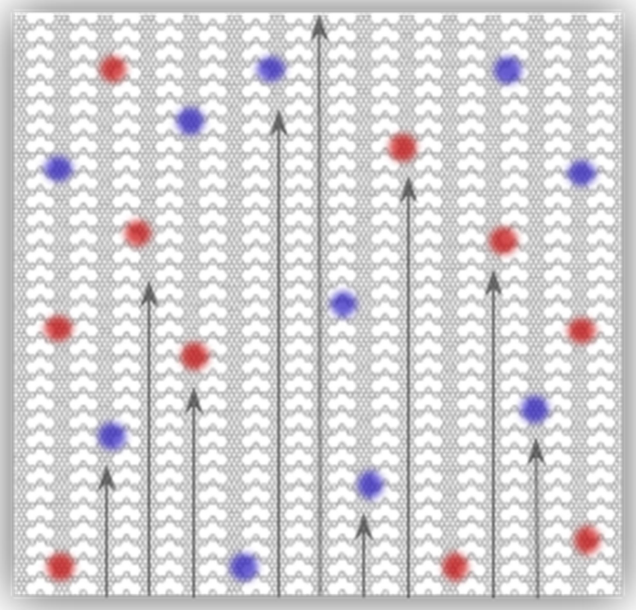